\documentclass[twocolumn,showpacs,preprintnumbers,amsmath,amssymb,superscriptaddress]{revtex4}

\usepackage{epsfig}
\usepackage{graphicx}
\usepackage{dcolumn}
\usepackage{bm}

\newcommand{\be}{\begin{equation}}
\newcommand{\ee}{\end{equation}}

\begin{document}

\preprint{APS/123-QED}

\title{Atom Interferometry with up to 24-Photon-Momentum-Transfer Beam Splitters}

\author{Holger M\"uller}
\email{holgerm@stanford.edu}
\author{Sheng-wey Chiow}
\author{Quan Long}
\altaffiliation{Now at Physics Department, Huazhong University of
Science and Technology, Wuhan 430074, Hubei, China.}
\author{Sven Herrmann}
\affiliation{Physics Department, Stanford University, 382 Via
Pueblo Mall, Stanford, California 94305, USA}
\author{Steven Chu}
\affiliation{Physics Department, Stanford University, 382 Via
Pueblo Mall, Stanford, California 94305, USA}
\affiliation{Lawrence Berkeley National Laboratory and Department
of Physics, University of California, Berkeley, 1 Cyclotron Road,
Berkeley, CA 94720.}

\date{\today}

\begin{abstract}
We present up to 24-photon Bragg diffraction as a beam splitter in
light-pulse atom interferometers to achieve the largest splitting
in momentum space so far. Relative to the 2-photon processes used
in the most sensitive present interferometers, these large
momentum transfer beam splitters increase the phase shift 12-fold
for Mach-Zehnder (MZ-) and 144-fold for Ramsey-Bord\'e (RB-)
geometries. We achieve a high visibility of the interference
fringes (up to 52\% for MZ or 36\% for RB) and long pulse
separation times that are possible only in atomic fountain setups.
As the atom's internal state is not changed, important systematic
effects can cancel.
\end{abstract}

\maketitle

Light-pulse atom interferometers are tools for experiments of
exquisite precision \cite{Pritchardreview}, such as tests of
general relativity \cite{LVGrav} and measurements of the
fine-structure constant $\alpha$ \cite{Weiss,Wicht,Biraben}, the
local gravitational acceleration $g$ \cite{Peters}, its gradient
\cite{Snaden98}, the Sagnac effect \cite{Gustavson}, or Newton's
gravitational constant \cite{Fixler}. Their sensitivity compares
favorably with the best competing methods. Substantial progress
would not only improve the above experiments, but also enable new
ones: For example, tests of gravity \cite{Jason}, quantum
electrodynamics \cite{Paris}, or the detection of gravitational
waves \cite{Dimopoulos}. Like in optical interferometers, the
sensitivity scales with the phase difference $\phi$ of the waves
in the interferometer arms. This can be increased via the arms'
momentum-space splitting. Conventionally, the splitting is
provided by the momentum of $2\hbar k$ that is transferred to the
atom by a two-photon process (where $k$ is the wavenumber).
Efforts to increase it have been limited to $6\hbar k$. They have
used momentum transfer by extra light pulses \cite{McGuirk,Gupta},
which might lead to additional systematic effects, or supersonic
atomic beams \cite{Giltner,Miffre}, whose inherently short
evolution time limits the sensitivity. Up to $140\hbar k$ have
been transferred by adiabatic transfer \cite{Weiss,Wicht}, but
this affects the common, not the relative momentum of the arms.
Here, we use multiphoton Bragg diffraction of atoms by an optical
lattice as a beam splitter. We achieve interferometry with a
momentum-space splitting of up to $24\hbar k$, the largest so far.
In some important applications \cite{Weiss,Wicht,Paris}, this
leads to a 144-fold increase in the phase. Moreover, Bragg
diffraction does not change the atom's internal state, so that
important systematic effects can cancel. This work thus allows for
substantial progress in both sensitivity and precision of atom
interferometry.

\begin{figure} \centering
\epsfig{file=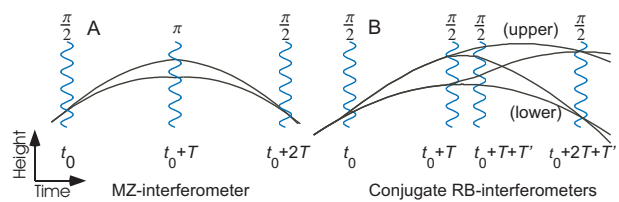,width=0.45\textwidth}
\caption{\label{schematic} A: MZI. ``$\pi/2$" pulses transfer
momentum with a probability of 1/2. They thus act as beam
splitters; ``$\pi$" pulses act as mirrors. B: Conjugate RBIs;
either is selected by the last $\pi/2$ pulse pair as described in
the text. Not shown are outputs of the third beam splitter, which
do not interfere.}
\end{figure}

In multiphoton Bragg diffraction, the atom coherently scatters
$2n$ photons from a pair of antiparallel laser beams, without
changing its internal state. The atom thereby acquires a kinetic
energy of $4n^2\hbar \omega_r$, where $\omega_r=\hbar k^2/(2M)$ is
the recoil frequency and $M$ the mass of the atom. Match with the
energy $n\hbar(\omega_1-\omega_2)$ lost by the laser field defines
the resonance condition for the difference frequency
$\omega_1-\omega_2$ of the beams. Bragg diffraction has been used
to transfer up to $16\hbar k$ \cite{Koolen}, but interferometry so
far has been limited to $6\hbar k$ \cite{Giltner,Miffre}, with up
to 26\% visibility of the interference fringes \cite{Miffre}.
These atomic beam setups are also limited by a relatively short
pulse separation time of $<1\,$ms.

The phase difference $\phi=\phi_F+\phi_I$ contains a contribution
of the atom's evolution between the beam splitters $\phi_F$, and
one of their interaction $\phi_I$. To discuss specifically the
effects of large momentum transfer beam splitters, it is useful to
consider Mach-Zehnder and Ramsey-Bord\'e interferometers (MZI and
RBI) separately. In MZIs (Fig. \ref{schematic} A), $\phi_F$
vanishes for constant $g$, but gravity causes a $\phi_I$ by
lowering the height at which the arms interact with the beam
splitters. If the momentum transferred by the beam splitter is
$2n\hbar k$, where $n$ is an integer, a MZI thus has a phase
difference of \cite{Pritchardreview,Peters} $\phi_{\rm
MZ}=n(2kgT^2-\phi_L)$, where $\phi_L=\phi_1-2\phi_2+\phi_3$ are
the phases $\phi_{1-3}$ of the laser fields at some reference
point. Here, multiphoton beam splitters lead to a linear increase
in phase. In RBIs, only one arm receives momentum from the beam
splitters (Fig. \ref{schematic} B). Thus, $\phi_F=2E_{\rm
kin}T/\hbar $ is nonzero due to the difference in kinetic energy
$E_{\rm kin}$. The same term, times minus two, enters $\phi_I$ due
to the modified locations at which the atoms interact. Summing up,
\cite{Wicht}
\begin{equation} \phi_{\rm RB}=\pm 8n^2\omega_r T+2n k
g(T+T')T+n\phi_L.
\end{equation}
The plus and minus signs are for the upper and lower
interferometer, respectively, and
$\phi_L=\phi_2-\phi_1-\phi_4+\phi_3$ is given by the phases
$\phi_{1-4}$ of the laser pulses. The recoil term in RBIs scales
quadratically with the momentum splitting. So far, the highest was
$4\hbar k$. It has been achieved by applying additional light
pulses \cite{Gupta}.

\begin{figure}
\epsfig{file=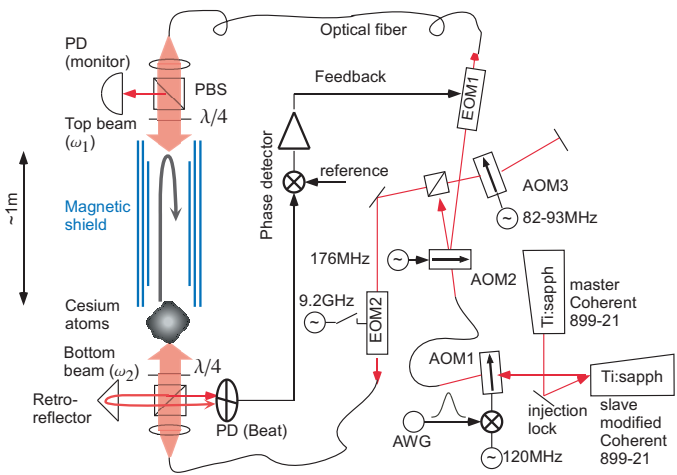,width=0.5\textwidth}
\caption{\label{setup} Setup (simplified). PD; photodetector. For
the injection lock, we use polarization spectroscopy
\cite{HanschCouillot}, which does not require modulation of the
laser light. AOM1 (Isomet 1206C) amplitude modulates, AOM2
(Crystal Technology 3200-124) splits and AOM3 (Isomet 1205C-1)
ramps the frequency of the beams. EOM2 (New Focus) generates $\sim
9.2$\,GHz sidebands for the velocity selection.}
\end{figure}

Our apparatus (Fig. \ref{setup}) loads $^{133}$Cs atoms from a
2-dimensional magneto-optical trap (2D-MOT; not shown) into a
3D-MOT. A moving optical molasses accelerates them upwards
(``launches") to a 1-m high, $0.9$-s ballistic trajectory, at a
temperature of $1.2-2\,\mu$K. Doppler-sensitive Raman transitions
driven by the top and bottom beams select $\sim 10^6$ atoms in the
6S$_{1/2}$, $F=3, m_F=0$ state with a narrowed vertical velocity
distribution of about $0.3\,v_r$ full width at half maximum
(FWHM). Here, $v_r=\hbar k/M\simeq 3.5\,$mm/s is the recoil
velocity for a wavelength of 852\,nm (the Cs D2 line).

High-powered laser beams are mandatory for driving high-order
multiphoton Bragg diffraction. On the one hand, the effective Rabi
frequency \cite{Losses,remark} $\Omega_{\rm eff} \approx
\Omega^n/[(8\omega_r)^{n-1}(n-1)!^2]$ is a very strong function of
the 2-photon Rabi frequency $\Omega$, i.e., of laser intensity. On
the other hand, beams of large radius are required to accommodate
the spread of the sample. To generate the required power, we use a
system of injection-locked Ti:sapphire lasers (Fig. \ref{setup}).
A first $\sim 1.2-$W Ti:sapphire laser is frequency stabilized
(``locked") to the 6S$_{1/2}$, $F=3\rightarrow 6P_{3/2}, F=4$
transition in a Cs vapor cell, with a blue detuning $\delta$ of
0-20\,GHz set by a microwave synthesizer. It injection locks a
second one, which has no intracavity
etalons or Brewster plate, and an output coupler with 10\% transmission (CVI part No. PR1-850-90-0537). 
Pumped with $17-19$\,W from a Coherent Innova 400 argon-ion laser,
it provides a single-frequency output power of up to 6\,W, about 2
times more than the strongest previously reported \cite{injlock}.

Acousto-optical modulators (AOMs) split the laser light into the
top and bottom beams and shape them into Gaussian pulses, defined
by arbitrary waveform generators (AWGs). Due to the free fall of
the atoms, the resonance condition for $\omega_1-\omega_2$ changes
at a rate of 23\,MHz/s, which we account for by continuously
ramping $\omega_1-\omega_2$ at a rate of $r$ by AOM3. The ramp
(provided by an Analog Devices AD9954 synthesizer) has a step size
of $\sim 0.01\,\mu$s, i.e., is essentially smooth even on the
time-scale of a single Bragg pulse.

Coherent Bragg diffraction at high order $n$ requires
proportionally lower optical wavefront distortions. To reduce
random aberrations, we minimize the number of optical surfaces.
The beams reach the experiment via 5-m long, single-mode,
polarization maintaining fibers and are collimated at a $1/e^2$
intensity radius of 8.6\,mm by a doublet lens featuring low
spherical aberration. Polarization is cleaned by 2" polarizing
beam splitter (PBS) cubes and converted to $\sigma^+-\sigma^+$ by
zero-order $\lambda/4$ retardation plates having a specified
$\lambda/20$ flatness.

For coherent high order Bragg diffraction, the beams also need to
have exceptionally low phase fluctuations between the top and
bottom beams. Therefore, we use a secondary phase lock
\cite{PLL,Paris}: The phase is measured by detecting the beat note
(Fig. \ref{setup}) and compared to an electronic reference.
Feedback is applied to EOM1 (New Focus 4002) via a fast
high-voltage amplifier \cite{HVAmp}. The $\sim 100\,$ns response
of this feedback loop allows us to re-lock at the beginning of
each pulse, within a time that is negligible compared to the pulse
length. The lock point relative to the reference will be the same
for each pulse, modulo $2\pi$, thus keeping the phase controlled
within and in between pulses.


The performance of Bragg beam splitters depends critically on the
choice of the duration, envelope function, and intensity of the
pulses \cite{Losses}. Our setup offers superior control of these.
Short pulses, with their large Fourier width, reduce the
sensitivity to the velocity spread of the atomic sample. However,
below an FWHM of $n^{1/6}/[\omega_r(n-1)]$ for Gaussian pulses,
losses into other diffraction orders become significant
\cite{Losses}. We use about $30-45\,\mu$s. At a detuning of
750\,MHz and a peak intensity of $0.5\,$W/cm$^2$ at the center of
each beam, $30\hbar k$ momentum transfer was achieved at $>50\%$
efficiency.

\begin{figure}
\epsfig{file=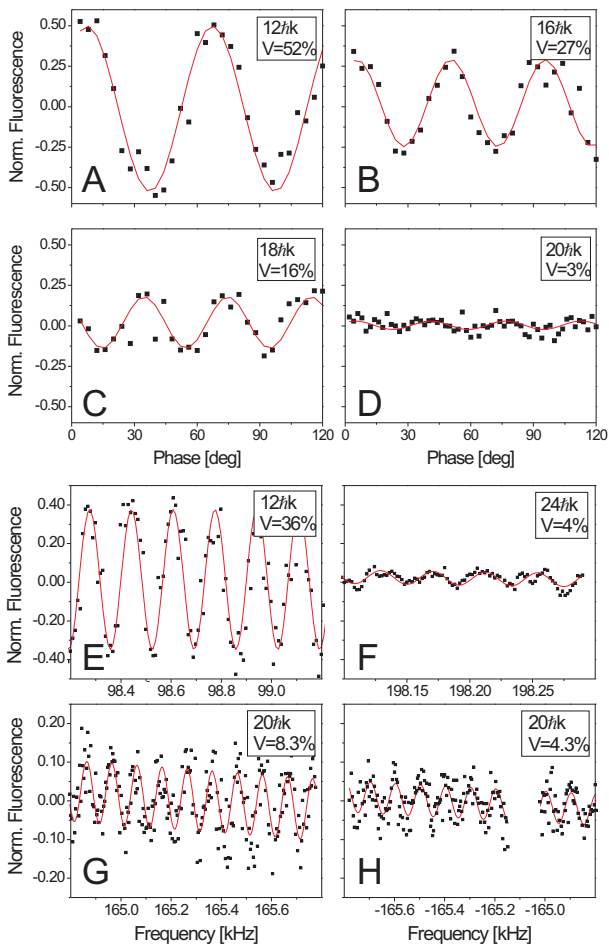, width=0.45\textwidth}
\caption{\label{MZGallery} A-D show MZ fringes with between 12 and
20$\hbar k$ momentum transfer; E and F are RB fringes with $12$
and $24\hbar k$. G and H show a conjugate $20\hbar k$ RB-pair.
Throughout, $T=1$\,ms, $T'=2\,$ms. Each data point is from a
single launch (that takes 2\,s), except for F, where 5-point
adjacent averaging was used. The lines represent a sinewave fit.}
\end{figure}

For MZ interferometry, we increase the detuning to $\delta=
4\,$GHz to further reduce single-photon processes. We generally
apply the first Bragg pulse about 100-200\,ms after launch, when
the thermal spread of the cloud is still negligible against the
radius of the Bragg beams. For $\leq 18\hbar k$ momentum transfer,
we achieve a $\pi$-pulse efficiency of 80-90\%. The fluorescence
$f_{1,2}$ of the two interferometer outputs is detected as they
pass a Hamamatsu R943-02 photomultiplier tube (located below the
magnetic shield in Fig. \ref{setup}). To take out fluctuations of
the initial atom number, we use the normalized fluorescence
$f=(f_1-f_2)/(f_1+f_2)$. We define the amplitude of a sinewave fit
of the measured fringes as the visibility $V$. Fig.
\ref{MZGallery}, A-D show fringes of MZIs with 12 to 20$\hbar k$
momentum transfer, measured by scanning the phase of the last beam
splitter. The period of the fringes is $2\pi/n$. Even at high
orders, excellent visibility is achieved, like $V=52\%$ at
$12\hbar k$. The strong decrease of $V$ at $20\hbar k$ is due to
insufficient laser intensity to drive higher-order multi-photon
transitions at $\delta=4\,$GHz.

For RBIs, we select the upper interferometer (Fig. \ref{schematic}
B), by shifting $\omega_1-\omega_2$ of the last pulse pair by
$\omega_u\simeq -8n\omega_r$, to meet the resonance condition for
addressing the upper interferometer arms. As
$\phi_L=rT^2+\omega_uT$ (where $r\simeq 2\pi\times 23\,$MHz/s is
the ramp rate), we measure the interference fringes by scanning
$\omega_u$, see Fig. \ref{MZGallery} E-H. Like for MZIs, we
achieve an excellent visibility, e.g., $V=36$\% at $12\hbar k$.
This is 72\% of the theoretical maximum, which is $V=50$\% because
each interferometer output overlaps with a fraction of the initial
population, which does not interfere (Fig. \ref{schematic} B). By
reducing $\delta$ to 3.3\,GHz, we can even increase the momentum
transfer to $24\hbar k$ and achieve $V=3.6\%$, which is still
useful. Note that the period of the fringes is $1/(nT)$. Since
additionally $\omega_u \propto n$, the resolution to which
$\omega_r$ can be measured increases by $n^2$, as expected.

Choosing an appropriate positive frequency shift $\omega_\ell$ for
the last $\pi/2$ pair forms a $20\hbar k$ lower RBI (Fig.
\ref{MZGallery} H). The contrast of this is somewhat reduced, as
background atoms that could not be diffracted by the Bragg pulses
overlap with one of the outputs. From a pair of conjugate RBIs,
which use the same $T$ and $T'$,
$\omega_r=(\omega_\ell-\omega_u)/(16 n)$ can be obtained
independent of $g, r, T,$ or $T'$. Here $\omega_\ell, \omega_u$
denote the values at the centers of the fringes.

\begin{figure}
\epsfig{file=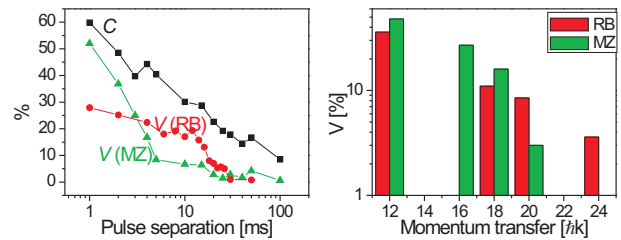, width=0.45\textwidth}
\caption{\label{ucontrast} Left: $C$ and $V$ (MZ) of $12\hbar k$
MZIs. $V$ (RB) shows larger visibility at longer times for a RBI
with vibration isolation (Minus K Technology Model 40BM1) for the
lower beam and beat optics (Fig. \ref{setup}). Right: Visibility
versus momentum transfer.}
\end{figure}

The visibility decreases for long $T$ and high $n$ (Fig.
\ref{ucontrast}). This may be ascribed to single-photon excitation
(large $n$ requires larger intensity), or thermal motion, which
removes the atoms from the center of the Bragg beams, or wavefront
distortions of the Bragg beams that smear out the phase over the
atomic sample. Moreover, in spite of the secondary phase lock,
phase noise enters in the atoms's inertial frame as a result of
the vibrations of the laboratory frame, in which the phase is
stabilized. At long $T$ (where the vibrations are harder to
isolate), the interferometers' phase thus becomes uncontrolled.
Thus, the normalized fluorescence will fluctuate with a contrast
$C=\sqrt{2}\sigma$, where $\sigma$ is the standard deviation, but
$V$ goes to zero. From the measurement of $C$ (Fig.
\ref{ucontrast}) we find that, while $V$ is reduced by vibrational
noise, interference is still taking place throughout to a pulse
separation time as long as $T=100$\,ms.

This work provides a tool for measurements of increased
sensitivity and accuracy. For example, the fine-structure constant
can be measured via the relation
$\alpha^2=(2R_\infty/c)(M/m_e)(h/M)$. The Rydberg constant
$R_\infty$ and the Cs to electron mass ratio $M/m_e$ are known to
precisions of 0.007 and 0.5 ppb, respectively \cite{CODATA}. A
non-interferometric measurement based on $\sim 450$ Bloch
oscillations \cite{Biraben} and an RBI with 30 additional $\pi$
pulses \cite{Wicht} both reach around $7$\,ppb in $\alpha$.
Replacing the beam splitters by 24 photon Bragg diffraction as
demonstrated here can increase the phase of the RBI by a factor of
$144$. For example, from data taken with $n=10$ (Fig.
\ref{MZGallery} G and H), we obtain $\omega_r=2\pi\times
2066.427(11)\,$Hz and $\alpha^{-1}=137.03653(35)$ [2.6\,ppm],
compatible at 1.5$\sigma$ with the accepted value. This
statistical uncertainty would be 260\,ppm if we had taken this
data [$T=1$\,ms, $V\sim (4-10)\%$] with $n=1$. While not being a
competitive measurement of $\alpha$, this clearly shows the power
of the method.

We expect to increase $T$ to 400\,ms with the help of the
vibration cancellation afforded by simultaneous conjugate RBIs
\cite{Paris,PLL}. 
As the sensitivity scales like $n^2T$, use of $n=12$ and
$T=400\,$ms offers a $\sim 500-$fold gain over the best previous
RBI, ($n=1, T=120$\,ms). Since the ultimate limit on the accuracy
will be systematics, such a large sensitivity is unnecessary;
operation without the 30 additional $\pi$ pulses would still be
sufficiently sensitive and help to reduce systematic effects by
simplifying the geometry to a basic RBI. This has an additional
benefit: the finite efficiency of the additional $\pi$ pulses
reduces the contrast to about 15\% in Ref. \cite{Wicht}. With
multiphoton Bragg diffraction, operation without additional $\pi$
pulses is possible, which can make up for the loss of contrast at
the highest Bragg diffraction orders (Figs.
\ref{MZGallery},\ref{ucontrast}).

Moreover, Bragg diffraction leaves the internal quantum states of
the atoms unchanged, so that systematic effects like the Zeeman
and Stark effects cancel out between the interferometer paths. (A
smaller contribution due to background field gradients remains.)
The thick Bragg beams with good wavefront quality used here reduce
other dominant systematic effects \cite{Weiss,Wicht,Biraben},
while the 80$\,\mu$rad phase noise of our laser system means that
the final accuracy can be reached within low integration time. For
other systematic effects and their suppression, see also
\cite{Paris,BPM,PLL}. A ppb-level measurement of $\alpha$ via
$\hbar/M$ could serve for testing of quantum electrodynamics by
comparison to $\alpha$ as derived (to 0.4\,ppb) from a measurement
of the electron's anomalous magnetic moment $g-2$
\cite{Gabrielse}. The influence of hadronic vacuum polarization
would be revealed, and bounds on low energy dark matter and a
possible internal structure of the electron could be established
via their hypothetical effect on $g-2$.

We have presented atom interferometers that use Bragg diffraction
for beam splitters that transfer up to 24$\hbar k$. Even with high
($12\hbar k$) momentum transfer, the visibility is comparable or
superior to typical interferometers based on 2-photon transitions.
Interference is observed up to a pulse separation of 100\,ms. Up
to $30\,\hbar k$ were transferred in a single diffraction. Factors
that lead to this progress include (i) improved understanding of
multiphoton Bragg diffraction \cite{Losses}, especially of the
influence of the pulse shape, (ii) a 6-W injection locked
Ti:sapphire laser system, (iii) good wavefront quality and large
diameter of the Bragg beams, and (iv) a secondary phase locked
loop to reduce phase noise. We have discussed applications for
precision atom interferometry. The potential of multiphoton Bragg
diffraction for atom interferometry is clear.

Many thanks to A. Peters and A. Senger for valuable help and to E.
Sarajlic and N. Gemelke for discussions. This work was supported
by the National Science Foundation under Grant No. 0400866, the
Multi-University Research Initiative, and the Air Force Office of
Scientific Research. H.M. and S.H. thank the Alexander von
Humboldt Foundation.

\end{document}